# Comment on "Counterintuitive Dispersion Violating Kramers-Kronig Relations in Gain Slabs"

In a recent Letter [1], Wang *et al.* considered the problem of the propagation of a plane wave through a slab with gain. The electromagnetic properties of the medium are described in the frequency domain in terms of its linear permittivity with a negative imaginary part. The authors considered the movement of singular points (SPs) of the reflection and transmission coefficients over the complex frequency plane. They argue that under certain conditions, SPs (poles) of the Fourier transforms of reflection and transmission coefficients of the slab may enter the upper half-plane of complex frequencies leading to the violation of the conventional Kramers-Kronig (CKK) relations. Then the authors come to a number of conclusions, including the existence of broadband "abnormal dispersion" in the gain slabs and an observable Hartman effect. The most unusual claim made by the authors is that the violation of the CKK relations should not lead to a violation of the causality principle. In this Comment, we refute these conclusions, which stem from the authors' use of the linear theory that is not applicable to systems having SPs in the upper half of the complex frequency plane.

The authors do not properly recognize that the move of SPs to the upper half-plane of complex frequencies indicates the onset of lasing. At this moment, the uniqueness of the solution of the Maxwell equations breaks down [2,3] and a nonzero solution in the *absence of incident field* arises. This nonzero solution to the linear Maxwell equations corresponds to a self-oscillating solution to the nonlinear Maxwell-Bloch equations, which is a direct indication of the transition to lasing appearing in real system as a Hopf bifurcation [4]. It is well known that if for a certain value of gain or the slab thickness a SP touches the real axis at some frequency $\omega_{SP}$, then the gain slab becomes a Fabry-Perot resonator with compensated losses [5]. At the frequency $\omega_{SP}$, the incident wave causes the field to diverge within the slab. Nonlinearity must be taken into account to obtain a bounded, physically reasonable solution. The mechanism for this nonlinearity is gain saturation connected together with the suppression of the inverse population by field.

The nonlinear interaction of the incident wave with this self-oscillation leads to either a stochastic or regular response depending on the strength of the incident field [4,6]. In any case, a transition from the linear to nonlinear regime occurs when an SP crosses the real axis is. At the transition point, not only the solutions but the gain medium itself begins to change. The solution inside the slab may be represented as a sum of travelling and standing waves. Above the lasing threshold, the standing optical wave inside the gain slab modulates the permittivity of the slab. As a result, the medium can no longer be treated as homogenous. This effect, known as the spatial hole burning [4], strongly alters scattering from such a slab.

The authors are correct that an appearance of SPs in the upper half-plane leads to the violation of the CKK relations. However they also claim that "although the existence of the SPs in the upper-half $\tilde{\omega}$ plane breaks the CKK relations, the causality of the gain slab system is always preserved". This unproved statement contradicts the Titchmarsh theorem, which states that causality and the validity of the CKK relations are equivalent (see, e.g., [7]). The reason for the coexistence of causality and a SP in the upper half-plane is inapplicability of the linear approach to the problem considered. As was mentioned above, the system response is nonlinear with a resonant behavior. As a result, one cannot Fourier transform the field in order to work in the frequency domain while introducing the CKK relations.

To summarize, the analysis of plane wave propagation through a gain medium based on a linear description, which leads to a number of unusual conclusions in Letter [1], is incorrect.

D. G. Baranov[1,2], A. A. Zyablovsky[1,2], A. V. Dorofeenko[1,2,3], A. P. Vinogradov[1,2,3], A. A. Lisyansky[4]

[1]Moscow Institute of Physics and Technology, 9 Institutsky per., Dolgoprudny, Moscow reg., Russia

[2]All-Russia Research Institute of Automatics, 22 Suschevskaya, Moscow, Russia

[3]Institute for Theoretical and Applied Electromagnetics RAS, 13 Izhorskaya, Moscow, Russia

[4]Department of Physics, Queens College of the City University of New York, Queens, NY 11367, USA